\documentclass[aps,pre,twocolumn,superscriptaddress]{revtex4-1}
\usepackage{amssymb}
\usepackage{amsmath,bm}
\usepackage{graphicx,color}
\usepackage{bbm}
\usepackage[bottom]{footmisc}
\usepackage{multirow}
\usepackage{xcolor}
\usepackage{xpatch} 
\makeatletter   
\usepackage{xpatch} 

\DeclareMathOperator{\Tr}{Tr}
\newcommand{\B}[1]{{\bm{#1}}}

\newcommand{\Lag}{\mathcal{L}}
\newcommand{\A}{\mathcal{A}}
\newcommand{\dif}{\mathrm{d}}
\newcommand{\rin}{r_\text{in}}
\newcommand{\rout}{r_\text{out}}
\newcommand{\C}[1]{{\mathcal{#1}}}

\begin{document}
\title{Elasticity, plasticity and screening in amorphous solids: a short review}
\author{Avanish Kumar}
\affiliation{Dept. of Chemical Physics, The Weizmann Institute of Science, Rehovot 76100, Israel}
\author{Itamar Procaccia} 
\affiliation{Dept. of Chemical Physics, The Weizmann Institute of Science, Rehovot 76100, Israel}
\affiliation{Sino-Europe Complex Science Center, School of Mathematics, North University of China, Shanxi, Taiyuan 030051, China.}

\begin{abstract}
The aim of this short review is to summarize the developing theory aimed at  describing the effect of plastic events in amorphous solids on its emergent mechanics. Experiments and simulations present anomalous mechanical response of amorphous solids where quadrupolar plastic events collectively induce distributed dipoles that are analogous to dislocations in crystalline solids. The novel theory is described, and a number of pertinent examples are provided, including the comparison of theoretical prediction to simulations or experiments.
\end{abstract}
\maketitle

\section{Introduction}

\begin{figure}[h!]
	\vskip 0.5 cm
	\includegraphics[width=0.95\linewidth]{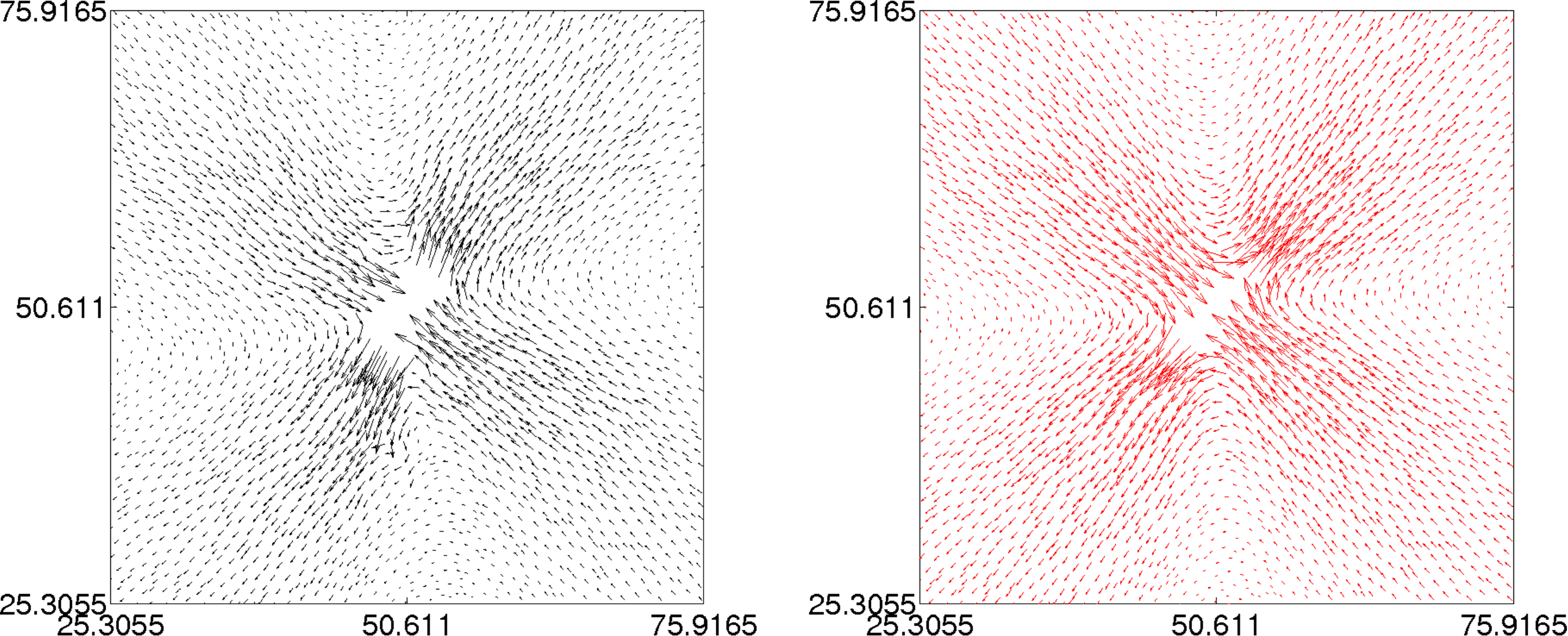}
	\caption{Left panel: Eample of a quadrupolar non-affine displacement field associated
		with a spontaneous plastic event \cite{12DHP} which
		is modeled by an Eshelby inclusion. Right
		panel: the displacement field associated with a single Eshelby
	 inclusion.}
	\label{quadrupole}
\end{figure}

Amorphous solids include materials like glasses, foams, colloids and granular media \cite{08Zal,98Ale}. These materials lack long-range order, and they do not possess a unique ``ground state". They can be cooled down to zero temperature, where they can reside in one of many available local equilibria, which under mechanical strains can easily exchange relative stability. Research in the last decade or two indicated that classical elasticity theory needs to be reconsidered for the treatment of amorphous solids. Contrary to perfect crystalline solids that can exhibit purely elastic response to strain, amorphous solids suffer from plastic responses, and these appear as quadrupolar Eshelby-like structures \cite{54Esh,99ML,06ML}, cf. Fig.~\ref{quadrupole}. 
The quadrupolar symmetry of localized plastic events stems from a purely geometric
conservation law, forbidding the formation of monopolar
and dipolar responses; quadrupolar plastic events are
the lowest order non-conserved multipoles  \cite{15KMS}. In amorphous solids any affine transformation results in non-affine irreversible responses, (i.e. plastic events) which appear (in the thermodynamic limit) at any infinitesimal deformation \cite{10KLPb,11HKLP}. Consequently both linear and nonlinear elasticity need to be reconsidered. Linear theory must change, the appearance of plastic events leads generically to screening and the emergence of typical length scales. In nonlinear elasticity where the stress is expanded in powers of the strain, the higher order elastic moduli have sample-to-sample fluctuations which can diverge upon increasing the system size \cite{11HKLP,16PRS,16DPSS,16BU,17DIPS}. The upshot of these findings is that plastic events cannot be neglected in formulating a mechanical theory of amorphous solids. 

For readers who are versed with the theory of dielectrics and ionic solutions \cite{11Fey}, it is useful to draw an analogy to the theory that is reviewed below. We will learn that quadrupoles in elasticity are analogous to dipoles in electrostatics, whereas dipoles in the former have effects similar to monopoles (charges) in the latter. Just as dipoles in dielectrics renormalize the dielectric constant (but do not change the nature of the theory), we will see that quadrupoles in elasticity do the same - they only renormalize the elastic moduli. In contradistinction, charges in electrostatics introduce screening, as is amply discussed in the context of Debye-H\"uckel theory of ionic solutions. We will see that dipoles in elasticity do the same, they bring about screening, an emergent length scale, and consequently a major modification of classical elasticity theory. In short, elasticity is "one-pole-up" compared to electrostatics. Below we provide a brief theoretical introduction to the subject and many example for the ramifications of this developing theory. 

\section{Anomalous elasticity}
\label{dilquad}

To derive equations for the displacement field $\B d (\B r)$ in the presence of quadrupoles, we start with an expression for the Lagrangian of the system, consistent with the underlying symmetries, presented up to quadratic order in the relevant fields. We present the theory of 3-dimensional system, but an equivalent 2-dimensional theory can be easily implied. 
The mechanical energy stored in the system stems from three main contributions \cite{13DHP}. First, the energetic cost associated with the bare imposed stress field $U_\text{el}$. Second is the interaction of the induced quadrupoles with the elastic background $U_\text{Q-el}$. Lastly, there is the self-interaction of the quadrupoles, reflecting their nucleation cost. Explicitly, 
$U = U_\text{el} + U_\text{Q-el} + U_\text{QQ}$. Defining the strain field
$u_{\alpha\beta} = [\partial_\alpha d_\beta +\partial_\beta d_\alpha]/2$, and the stress field $\sigma_{\alpha\beta}= A^{\alpha\beta\gamma\delta} u_{\gamma\delta}$, we write:
\begin{eqnarray}
	&&	U_\text{el} =\!\!\int \dif^3 x \frac{1}{2} A^{\alpha\beta\gamma\delta} u_{\alpha\beta}u_{\gamma\delta},~
		U_\text{Q-el} =  \int \dif^3 x \Gamma^{\alpha\beta}_{\gamma\delta} u_{\alpha\beta}Q^{\gamma\delta}\nonumber \\
		&&U_\text{QQ} =\int \dif^3 x \mathcal{F} \left(Q^{\alpha\beta},\partial_\beta Q^{\alpha\beta}\cdots\right)\ . 
	\label{eq:energydecomp}
\end{eqnarray}
Here $ A^{\alpha\beta\gamma\delta}$ is the standard tensor of elastic moduli, and  $\B \Gamma$ is an appropriate coupling tensor, that eventually renormalizes the standard moduli.  $\mathcal{F}$ represents the energy cost of the induced plastic quadrupoles, including their first and higher gradient terms. When the gradient terms are important they lead to screening, and see below for details. 
\subsection{The quasi-elastic regime}
In the dilute quadrupoles limit, corresponding to large
energetic cost for nucleating dipoles, quadrupoles vary
slowly in space to avoid effective dipoles, hence \cite{21LMMPRS,23CMP}
$\C F=\mathcal{F} \left(Q^{\alpha\beta}\right) $:
\begin{equation}
	\mathcal{F} \left(Q^{\alpha\beta}\right) = \frac{1}{2}\Lambda_{\alpha\beta\gamma\delta} Q^{\alpha\beta}Q^{\gamma\delta}	
\end{equation}
Upon minimizing $U$ with respect to the fundamental fields $d$ and $Q$, using  \eqref{eq:energydecomp}, we find  \cite{21LMMPRS,23CMP}
 a linear screening relation (analogous to the linear relation between electric field and induced polarization in dielectric materials \cite{49Fro})
\begin{equation}
	Q^{\alpha\beta} = - \Lambda^{\alpha\beta\mu\nu}  \Gamma_{\mu\nu}^{\gamma\delta} u_{\gamma\delta} \equiv  - \tilde{\Lambda}^{\alpha\beta\gamma\delta} u_{\gamma\delta} \ ,
	\label{eq:Screeningrelation}
\end{equation}
where $\Lambda^{\alpha\beta\mu\nu}$ is the inverse of $\Lambda_{\alpha\beta\mu\nu}$. 
The second result that one finds is
\begin{equation}
	\partial_\alpha  \tilde{\sigma}^{\alpha\beta} = 0 \ ,
	\label{eq:Equilibrium}
\end{equation}
where $\tilde \sigma^{\alpha\beta} \equiv \sigma^{\alpha\beta}+ \Gamma^{\alpha\beta}_{\gamma\delta} Q^{\gamma\delta}$.
We see that the re-normalization of the quadrupole-quadrupole interactions results with a linear constitutive relation between inducing stress and induced quadrupoles which then renormalizes the elastic tensor \cite{20NWRZBC}. This is the analog of the situation in dielectrics, where the dielectric constant is renormalized
by the induced dipoles \cite{49Fro}. Explicitly, the tensor of moduli is renormalized as follows:
\begin{equation}
	\tilde A^{\mu\nu\rho\sigma}\equiv A^{\mu\nu\rho\sigma}+\Lambda_{\alpha\beta\gamma\delta}\tilde{\Lambda}^{\alpha\beta\mu\nu}\tilde{\Lambda}^{\gamma\delta\rho\sigma}-2\Gamma_{\mu\nu}^{\gamma\delta}\tilde{\Lambda}^{\gamma\delta\rho\sigma} \ .
\end{equation}
Using this renormalized tensor, the Lagrangian in the quasi-elastic regime
can be written again in the form 
\begin{equation}
{\cal L}\equiv	\frac{1}{2}\tilde A^{\alpha\beta\gamma\delta} u_{\alpha\beta}u_{\gamma\delta} \ ,
\label{renL}
\end{equation}
leaving the form of the theory unchanged. 
\subsection{The screening regime}
\label{highdensity}
At high quadrupole densities one cannot neglect the gradient terms. Upon denoting the elastic tensor dressed by the induced quadrupolar terms (as discussed in the last subsection) by $\tilde{\A}$, we now consider the gradient terms in the function $\mathcal{F} \left(Q^{\alpha\beta},\partial_\beta Q^{\alpha\beta}\cdots\right)$, $U=\int d^3x \Lag$ and the Lagrangian $\Lag$ reads \cite{21LMMPRS,23CMP}:
\begin{equation}
	\begin{split}
		\Lag &=  \frac{1}{2} \tilde{A}^{\mu\nu\rho\sigma} u_{\mu\nu}u_{\rho\sigma} + 
		\frac{1}{2} \Lambda_{\alpha\beta} \partial_\mu Q^{\mu\alpha}  \partial_\nu Q^{\nu\beta}
		+ \Gamma_{\alpha}^{\,\,\beta} \partial_\mu Q^{\mu\alpha} d_{\beta} \ .
	\end{split}
	\label{lagdip}
\end{equation} 
the last term here results from $U_{Q-el}$ in \eqref{eq:energydecomp} by integration by parts. Note that the quadrupole-quadrupole terms were not included since the renormalization of the moduli is already taken into account.

Denoting the gradients on the quadrupoles as effectively induced dipoles  $P^\alpha \equiv \partial_\beta Q^{\alpha\beta}$, and minimizing with respect to the fundamental fields $d$ and $Q$ we find 
\begin{eqnarray}
	P^{\alpha} &=& -  \Lambda^{\alpha\beta}  \Gamma_{\beta}^{\gamma} d_{\gamma}\ ,\nonumber\\ 
	\partial_\alpha \sigma^{\alpha\beta} &=& \Gamma_{\alpha}^{\beta} P^{\alpha} = -\Gamma_{\alpha}^{\beta}  \Lambda^{\alpha\mu}  \Gamma_{\mu}^{\gamma} d_{\gamma} 
	 \ .
	\label{eq:ScreeningEquilibrium2}
\end{eqnarray}
We see that the displacement field acts as a screening source in the equilibrium equation. We  now rewrite this equation by substituting the stress in terms of strain, and the strain in terms of the displacement. In isotropic and homogeneous materials in Cartesian geometry the coupling tensors have the following forms
\begin{eqnarray}
		&&\tilde A^{\alpha\beta\gamma\delta} = \lambda \delta^{\alpha\beta} \delta^{\gamma\delta} + \mu \left(\delta^{\alpha\gamma}\delta^{\beta\delta}+\delta^{\alpha\delta} \delta^{\beta\gamma}\right)\nonumber \\
	&&	\Gamma_{\alpha}^{\beta} = \mu_1\delta_{\alpha}^{\beta} \ , \quad
	\Lambda_{\alpha\beta} =\mu_2 \delta_{\alpha\beta} \ ,
\end{eqnarray}
with $\mu_1$, $\mu_2$ being scalar coefficients.
Direct substitution yields
\begin{equation}
	\Delta \mathbf{d} + \left(1+\frac{\lambda}{\mu}\right) \nabla \left(\nabla\cdot \mathbf{d}\right) = -\frac{\mu_1^2}{\mu_2 \mu } \mathbf{d} \ .
	\label{final}
\end{equation}
The screening effect is negligible when $\frac{\mu_1^2}{\mu_2 \mu } \ll 1$. 
Unlike the quadrupole screening, dipole screening leads to a qualitatively new equation. The appearance of the displacement field $\B d$ without gradients represents a breaking of translational symmetry. The parameter $\kappa$, defined via $\kappa^2\equiv \frac{\mu_1^2}{\mu_2 \mu }$, is an inverse scale. The appearance of a scale heralds the breakdown of classical elasticity theory, leading to qualitatively new mechanical responses as is shown in the next section in which                                                                                                                                                                                                                                                                                                                                                                                                                                                                                                                                                                                                                                                                                                                                                                                                                                                                                                                                                                                                                                                                                                                                                                                                                                                                                                                                                                                                                                                                                                                                                                                                                                                                                                                                                                                                                                                                                                                                                                                                                                                                                                                                                                      we provide examples for the consequences of Eq.~(\ref{final}), both in two and three dimensions. The equation was shown to be valid in both cases. 
Ideas of how to predict a-priori the numerical value of the emergent inverse scale $\kappa$ can be found in \cite{23JPS} and in Subsect. ~\ref{transition} below. 
\section{Examples}

In this section we present a few examples of the consequences of Eq.~(\ref{final}). In all of these examples either the external strain or the plastic responses themselves will impose nonuniform quadrupolar field that results in dipole screening.
\subsection{Radial inflation in 2-dimensions}
\label{radinf}

For the purpose of analytic solutions of Eq.~(\ref{final}), the easiest examples are radial and spherical inflations of an inner boundary in a circle or a ball of much large outer boundary, which is filled up with an amorphous solids. In two dimensions we consider an annulus of radii $\rin$ and $\rout$ with an imposed displacement $\mathbf{d}(\rin) = d_0 \hat{r}$ and $\mathbf{d}(\rout) = 0$. The polar symmetry of the problem implies that $\mathbf{d}(r) =d_r(r) \hat{r}$, and Eq.~(\ref{final}) becomes,
In polar coordinates \cite{21LMMPRS,23CMP}, 
\begin{equation}
	d_r'' +\frac{1}{r} d_r' +(\kappa^2 -\frac{1}{r^2})d_r=0 \ .
	\label{radial2d}
\end{equation}
This is the Bessel equation. A solution 
of this equation satisfying $d_r(r_{\rm in})=d_0$, $d_r(r_{\rm out})=0$ reads
\begin{equation}
	d_r(r)  = d_0 \frac{ Y_1(r \, \kappa ) J_1(r_\text{out} \kappa )-J_1(r \, \kappa ) Y_1(r_\text{out} \kappa )}{Y_1( r_\text{in} \kappa ) J_1(r_\text{out} \kappa )-J_1(r_\text{in} \kappa ) Y_1(r_\text{out} \kappa )} \ .
	\label{amazing}
\end{equation}
Here $J_1$ and $Y_1$ are the Bessel functions of the first and second kind respectively.
The reader should note that the classical elastic solution is obtained from this Eq.~(\ref{radial2d}) using $\kappa=0$. The solution in that case reads
\begin{equation}
	{d}_r(r)=d_0 \frac{r^2 - \rout^2}{\rin^2 - \rout^2}\frac{\rin}{r} \ .
	\label{renelas}
\end{equation}
The nature of this solution is very different from the one presented in Eq.~(\ref{amazing}) The elastic solution stays positive, it decays in the bulk like $1/r$ and falls to zero at the outer boundary. The solutions of Eq.~(\ref{amazing}) can show oscillations, maxima, negative regions and what not,  determined by the  ratio between $r_\text{out}$ and $\kappa^{-1}$.
Probably the most surprising effect of the screening implied by the inverse scale $\kappa$ is that particle can move inward even though the inflation of the inner boundary points outward. 
\subsubsection{Verification by experiments and simulations} 
\begin{figure}
	\includegraphics[width=1.0\linewidth]{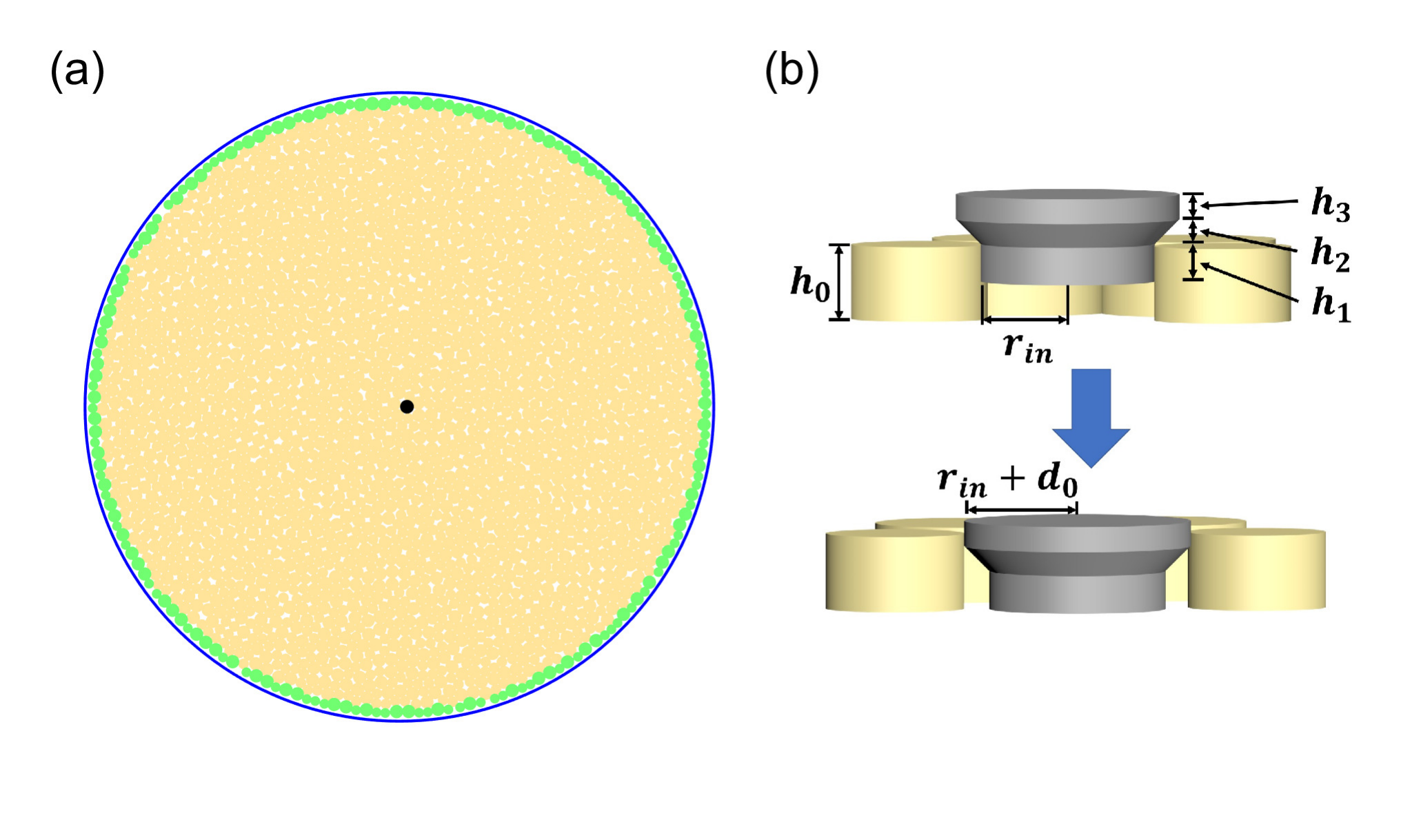}
		\caption{Panel (a): Top view of the experimental system. The blue line marks the position of the circular boundary. The green  
		layer represent the photo-elastic disks at the circumference, which are used as pressure sensors. In the yellow area, 
		bi-disperse disks are filled in, see Ref.~\cite{22MMPRSZ} for details. The black dot in the 
		center represents the conical shaped pusher used to achieve the inflation. Panel (b) A diagram of the inflation process, see Ref.~\cite{22MMPRSZ} for details.}
	\label{expt}
\end{figure}

\begin{figure}
	\includegraphics[width=1.1\linewidth]{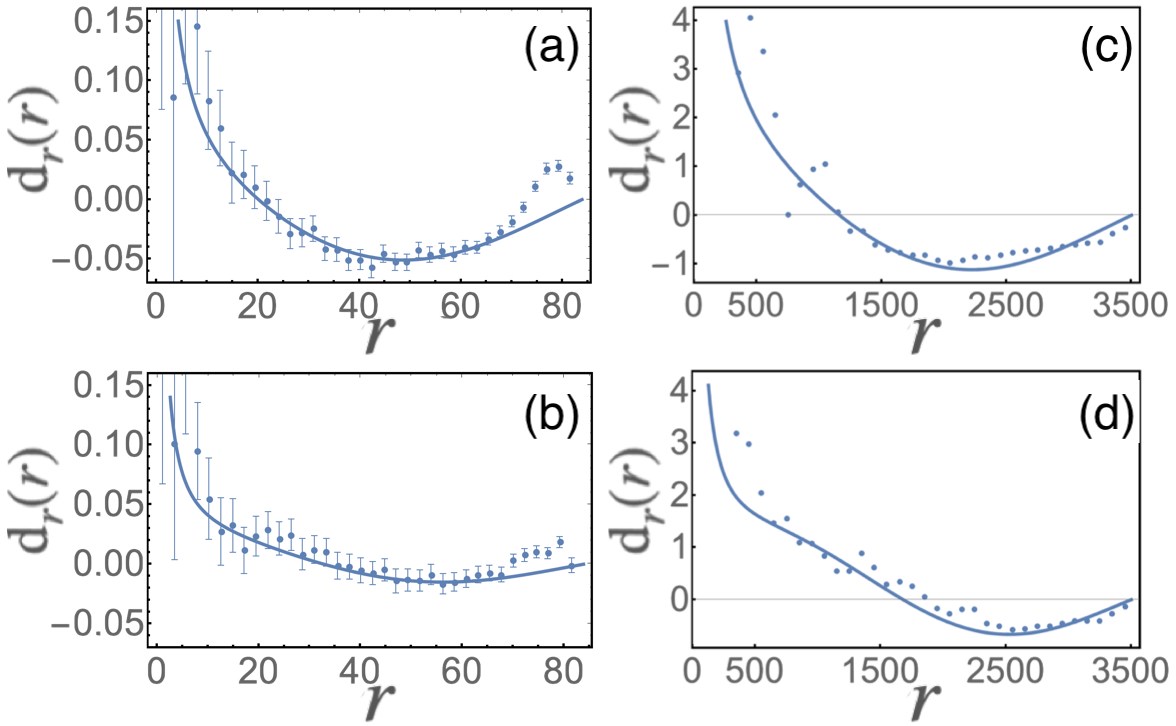}
	\caption{Radial component of the Displacement field induced by the inflation of one disk at the center of the box. Panels (a) and (b) simulation  results at dimensionless pressure $\tilde P=2.25\times 10^{-6}$, $\phi\approx 0.872$ with inflation of 10\%. The continuous lines are the analytic solution Eq.~(\ref{renelas}). Here $r_{\rm in} =1.14$ and  $r_{\rm out} =83.82$. $d_0=0.7$ and 1 in panels (a) and (b) respectively. Panels (c) and (d): experimental results at pressure $\tilde P\approx 10^{-3}$ with inflation of 10\%. Here $r_{\rm in} =140$ and  $r_{\rm out} =3500$, $d_0=19.1$ and 18.3 respectively.}
	\label{expts}
\end{figure}

A typical experiment to test the predictions of the theory was described in full detail in Ref.~\cite{22MMPRSZ}, and it is sketched in Fig.~\ref{expt}.
In Fig.~\ref{expts} we show typical examples of the radial component of the displacement fields as measured in this experiment and in numerical simulations. These data were obtained by performing angle averaging on the raw displacement field data, leaving a function of $r$ only. The radial component of the displacement field differs qualitatively from the linear elastic prediction Eq.~(\ref{renelas}), and is in good agreement with the screening theory. 
\vskip 0.5 cm

\subsection{Spherical inflation in 3-dimensions}
\begin{figure}
	\includegraphics[width=0.9\linewidth]{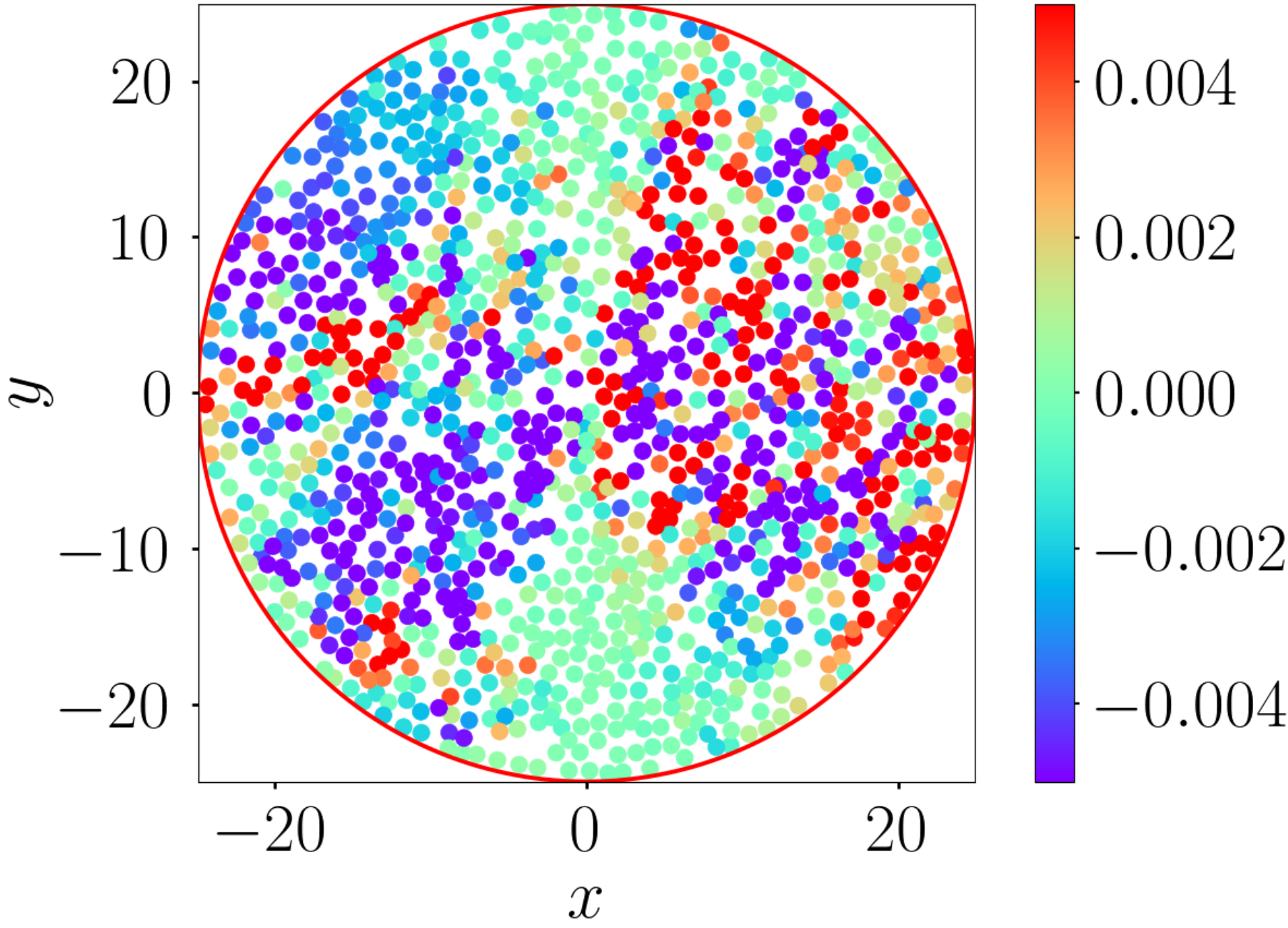}
	\includegraphics[width=0.9\linewidth]{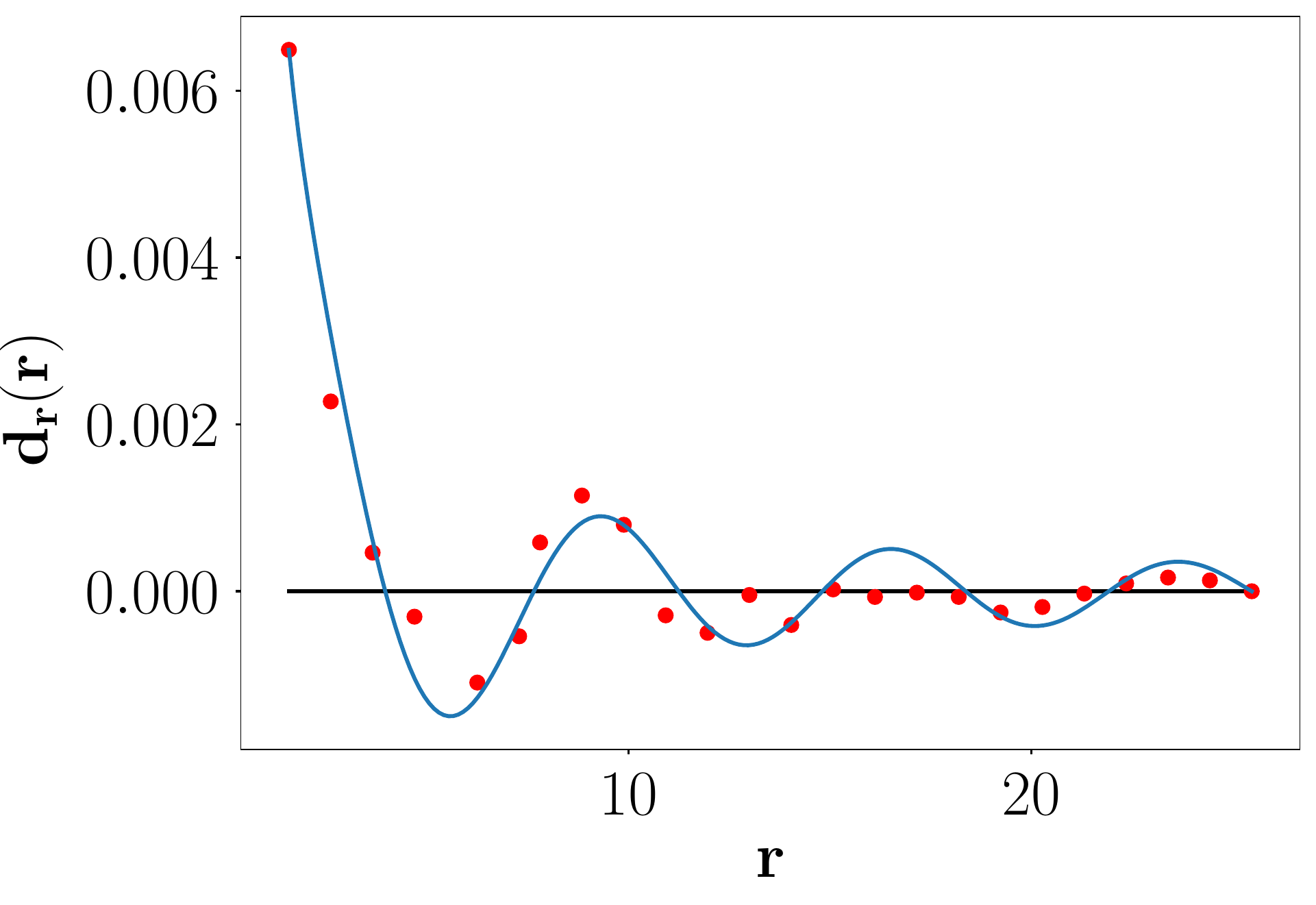}
	\caption{Radial displacement field with 30\% inflation, $\phi=0.649$. Panel a: Radial displacement field in a planar cross section of the three-dimensional sphere at $z=0$. Panel b: Comparison of the spherical averaged displacement field with $K_n=2000$ to the theory Eq.~(\ref{amazing}), using $\kappa=0.669$.}
		\label{3D}
	\end{figure}

In 3-dimension we do not have at this time experiments, but we did perform extensive simulations. The simulations employed binary assemblies of balls of two sized, interacting via Hertzian binary forces. Details of the simulations can be found in \cite{23CMP}. In Fig.~\ref{3D} we present typical results of a simulation that begins with a configuration of $N =42876 (35^3+1)$ bi-disperse disks placed randomly in a spherical volume with a radius, $r_{\rm out}=25.48$ in units of radius of our smaller sphere. The inner sphere that is inflated starts with $r_{\rm in}=1.56$ and $d_0=0.0065$. The oscillations seen are in strong support of the anomalous theory presented in the last section. 

\subsection{The Eshelby problem in amorphous solids}

A particularly interesting example for the study of mechanical responses of amorphous solids to strain is the Eshelby problem. The ``Eshelby problem" consists of computing the displacement field resulting from cutting out a circle from an elastic sheet, deforming it into an ellipse and pushing it back \cite{54Esh}. Surprisingly, it turned out that this seemingly artificial problem is intimately connected to the physics of plastic events in strained amorphous solids, cf. Fig.~7 in \cite{06ML}. Accordingly, the ``Eshelby problem" has become popular and a frequently employed theory to discuss the redistribution of stress after plastic events. In particular, the Eshelby kernel was often used in the context  of ``elasto-plastic" models which purport to describe the mechanical response of amorphous solids to external strains, up to mechanical yield by shear banding  \cite{98HL,06Sol,17NFMB}. 
\begin{figure}
	\vskip 0.5 cm
	\includegraphics[width=0.40\linewidth]{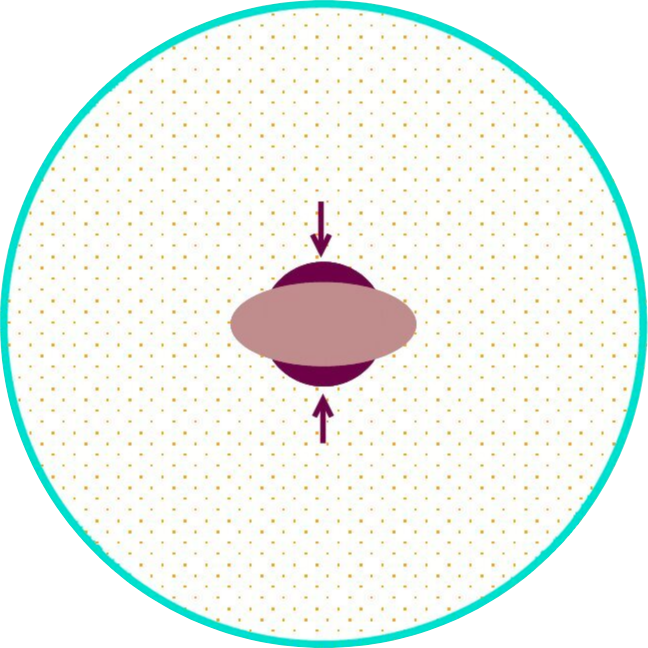}
	\caption{The geometry used: amorphous solid is contained between the outer circle of radius $r_{\rm out}$ and an inner circle of radius $r_{\rm in}$, which is then distorted to an ellipse of of the same area. We are interested in the displacement field as a result of this distortion. }
	\label{geometry}
\end{figure}
We considered the Eshelby problem in circular geometry, see Fig.~\ref{geometry} \cite{23HKPR}. Initially the amorphous material is confined between an inner circular cavity of radius $r_{\rm in}$ and an outer circle of radius $r_{\rm out}$. The inner circle is deformed to an ellipse of the same area, with major semi-axis $a$ and minor semi-axis $b$, such that $ab =r_{\rm in}^2$. The boundary of the ellipse is now
$
	\frac{x^2}{a^2} + 	\frac{y^2}{b^2} =1\ . 
$
Defining $\delta\equiv a/r_{\rm in}$, the boundary of the ellipse $r(\theta)$ is traced by 
\begin{equation}
	r(\theta) = \frac{r_{\rm in}}{\sqrt{\cos^2(\theta)/\delta^2 + \delta^2\sin^2 \theta}}\ ,	
\end{equation}
where $\theta=\arctan(y/x)$. We are interested in the displacement field that responds to the change from circle to ellipse, with radial component $d_r(r,\theta)\hat r$ and transverse component $d_\theta (r,\theta)\hat \theta$, where $\hat r$ and $\hat \theta$ are unit vectors in the radial and the transverse directions. 
The boundary condition on the outer circle are $\B d(r=r_{\rm out})=0$ , and on the ellipse 
$d_r (\theta) = r(\theta)-r_{\rm in}$ and $d_\theta (\theta)=0$. 
We followed in the footsteps of Eshelby, solving analytically the Eshelby problem for a small distortion of the ellipse, i.e. $\delta=1+\epsilon+\cdots$.  Then on the ellipse 
\begin{equation}
	d_r(\theta)	= r_{\rm in}\epsilon \cos(2\theta) \ , \quad d_\theta(\theta)=0 \ . 
	\label{bc}
\end{equation}

\begin{figure}
	\includegraphics[width=0.83\linewidth]{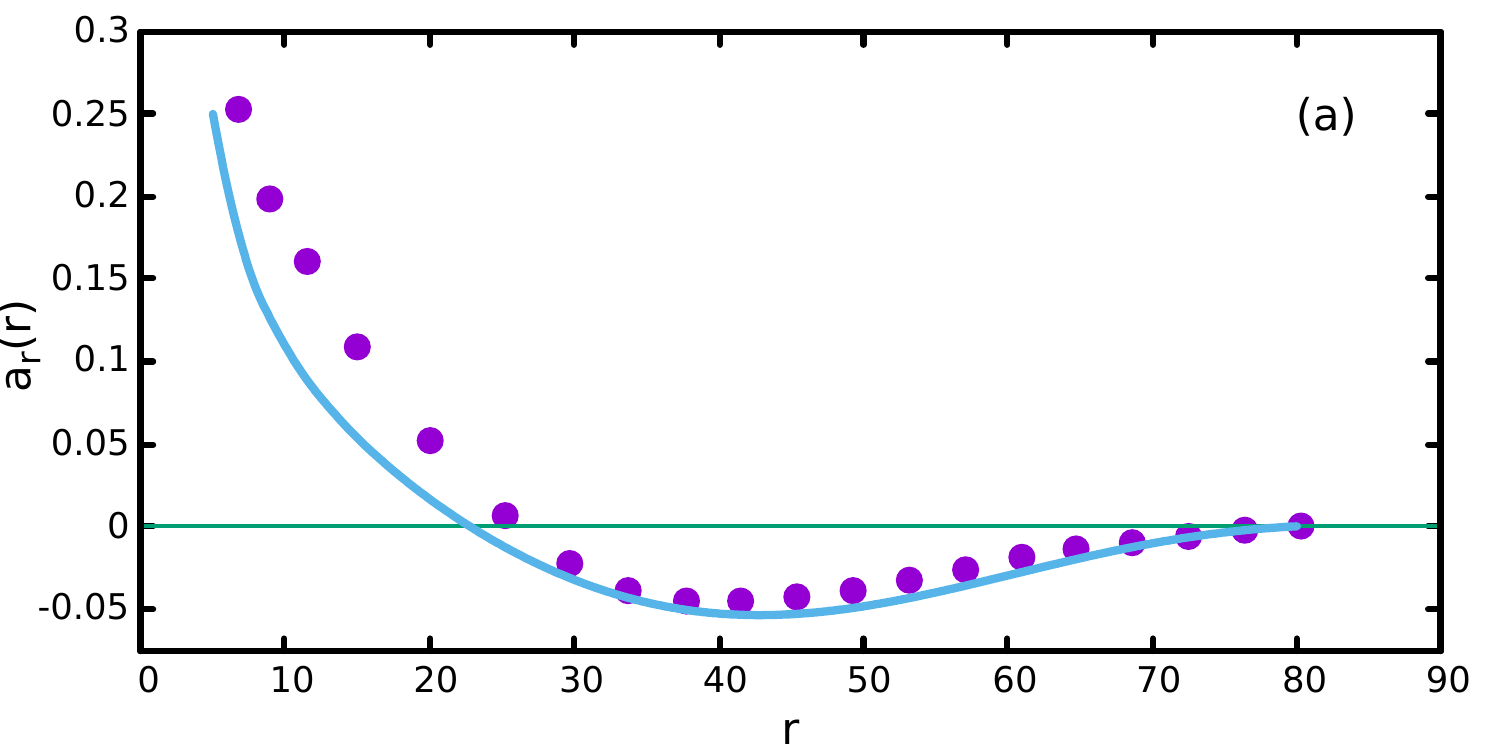}
	\includegraphics[width=0.83\linewidth]{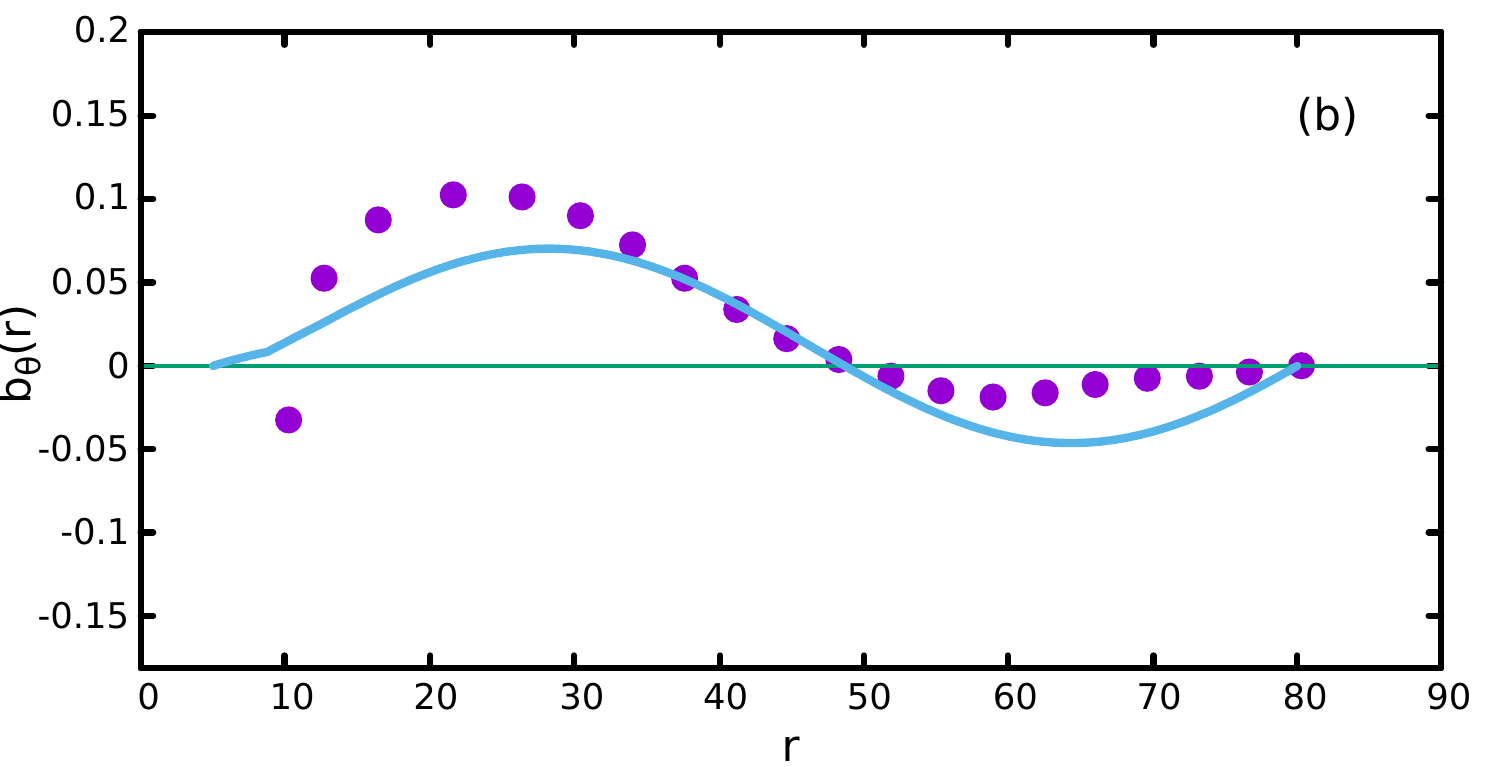}
	\caption{Comparison of the functions  $a_r(r)$ and $b_\theta(r)$ from the analytic solution with $\kappa=0.25$ and the angle average of the numerical simulations. Here $\mu=6.21$ and $\lambda=60$.}
	\label{kap}
\end{figure}

Equation (\ref{final}) can be solved analytically subject to the boundary conditions Eq.~(\ref{bc}). The solution is described in Ref.~\cite{23HKPR}, with the final result expressed in terms of the radial and tangential components
\begin{eqnarray}
	d_{r}(r,\theta) &=&   a_{r}(r)\cos(2\theta) \nonumber \\
	d_{\theta}(r,\theta)& =&  b_{\theta}(r)\sin(2\theta)  \ .
	\label{defab}
\end{eqnarray}
 Simulations employing Lennard-Jones poly-dispersed glasses were performed, and results were favorably compared to the analytic solutions, cf. Fig.~\ref{kap}. 

\subsection{Pure shear}

A priori simple or pure shear are not ideal protocols for the study of anomalous elasticity, since they are imposing a uniform strain, while dipoles appear due to nonuniform quadrupolar fields. Gradients are essential for screening. Nevertheless in amorphous solids simple and pure shear are immediately punctuated by plastic events, and these impose nonuniform strain and therefore also screening. 

 To demonstrate the issues we chose as our example frictional granular matter in a 2-dimensional square box of size $Lx_0\times Ly_0$, and applied pure shear by contracting the $x$-direction and pulling along the $y$-direction. Details of the simulations can be found in Ref.~\cite{23MMPR}. We measured the instantaneous pressure $P$ and the accumulated {\em affine} strain 
\begin{equation}
	u_{\rm aff}\equiv \frac{1}{2}\big(\frac{Lx_0-Lx}{Lx_0}+\frac{Ly-Ly_0}{Ly_0}\big) \ ,
\end{equation} 
where Lx and Ly are the instantaneous box-lengths along x and y directions respectively.  As is usual in such simulations, we observe intervals of increase in stress when the strain increases, interrupted by sharp drops in stress due to plastic events. These are the events that we focus on next.  Denoting the positions of our $N$ disks before and after the event as $\B r_i^a$ and $\B r_i^b$ respectively, we compute  the displacement field as $\B d_i\equiv \B r_i^a-\B r_i^b$. Next we compute the total strain field as
$u_{ij} = 0.5(\nabla_i d_j  + \nabla_j  d_i )$.
The non-affine strain $\B u^q$ is obtained by subtracting the affine strain  generated in the last step from $u_{\rm tot}$, 
\begin{eqnarray}
	&&u^q_{11}\equiv u_{11} - \frac{1}{2}\big(\frac{Lx^b-Lx^a}{Lx^b}\Big)\ ,
	u^q_{22}\equiv u_{22}-\frac{1}{2}\big(\frac{Ly^a-Ly^b}{Ly^b}\big) \ , \nonumber\\
	&&u^q_{12}\equiv u_{12}\ , \quad u^q_{21}\equiv u_{21} \ .
\end{eqnarray}
where again `a' and `b' refer to after and before. Having the non-affine strain we decompose it into its trace and its traceless components (cf. Ref.~\cite{15MSK} page 6):
\begin{equation}
	\B	u^q = m\B I +Q \B u^{ts} \ ,
\end{equation}
where $\B I$ is the identity tensor and $\B u^{ts}$ a traceless symmetric tensor. In the last equation $m=0.5 \Tr u_q$ and 
\begin{equation}
	Q^2 =  (u^{ts}_{11})^2  +  (u^{ts}_{22})^2 \ .
	\label{defQ}
\end{equation}
The quadrupolar charge $Q$ is obtained as the square root, and its orientation is computed from \cite{15MSK}:
\begin{equation}
	\Theta = 0.5 \arctan ((u^{ts}_{12})/(u^{ts}_{11}) ) \ .
	\label{deftheta}
\end{equation}

\begin{figure}
	\includegraphics[width=0.8\linewidth]{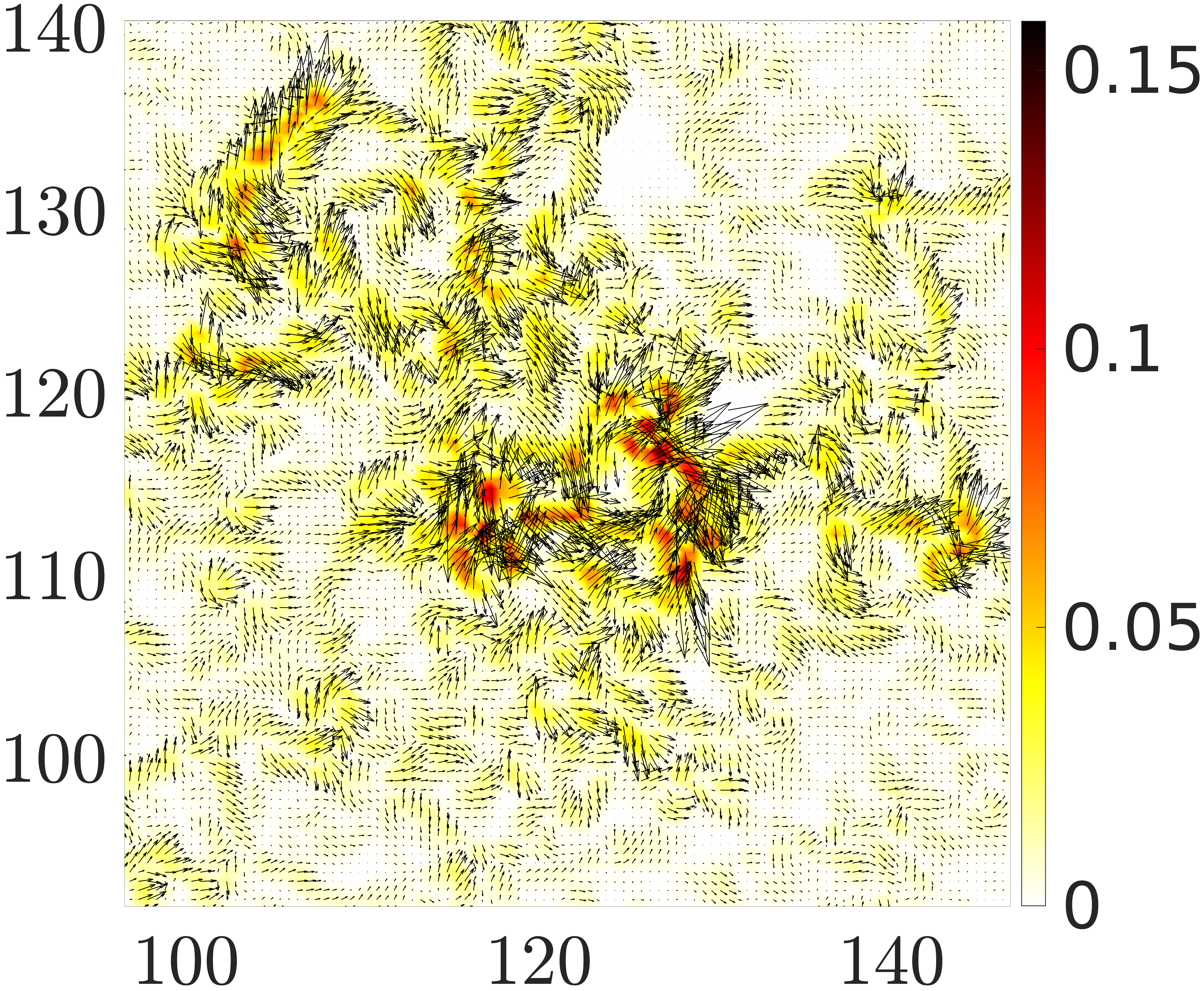}
	\caption{Heat map of the quadrupolar field for a part of our system after a plastic event at a pressure
		$P_0=4.5$. The darker region indicate high values of $Q$ cf. Eq.~(\ref{defQ}), and light region low values. }
	\label{Q}
\end{figure}

\begin{figure}
	\includegraphics[width=0.8\linewidth]{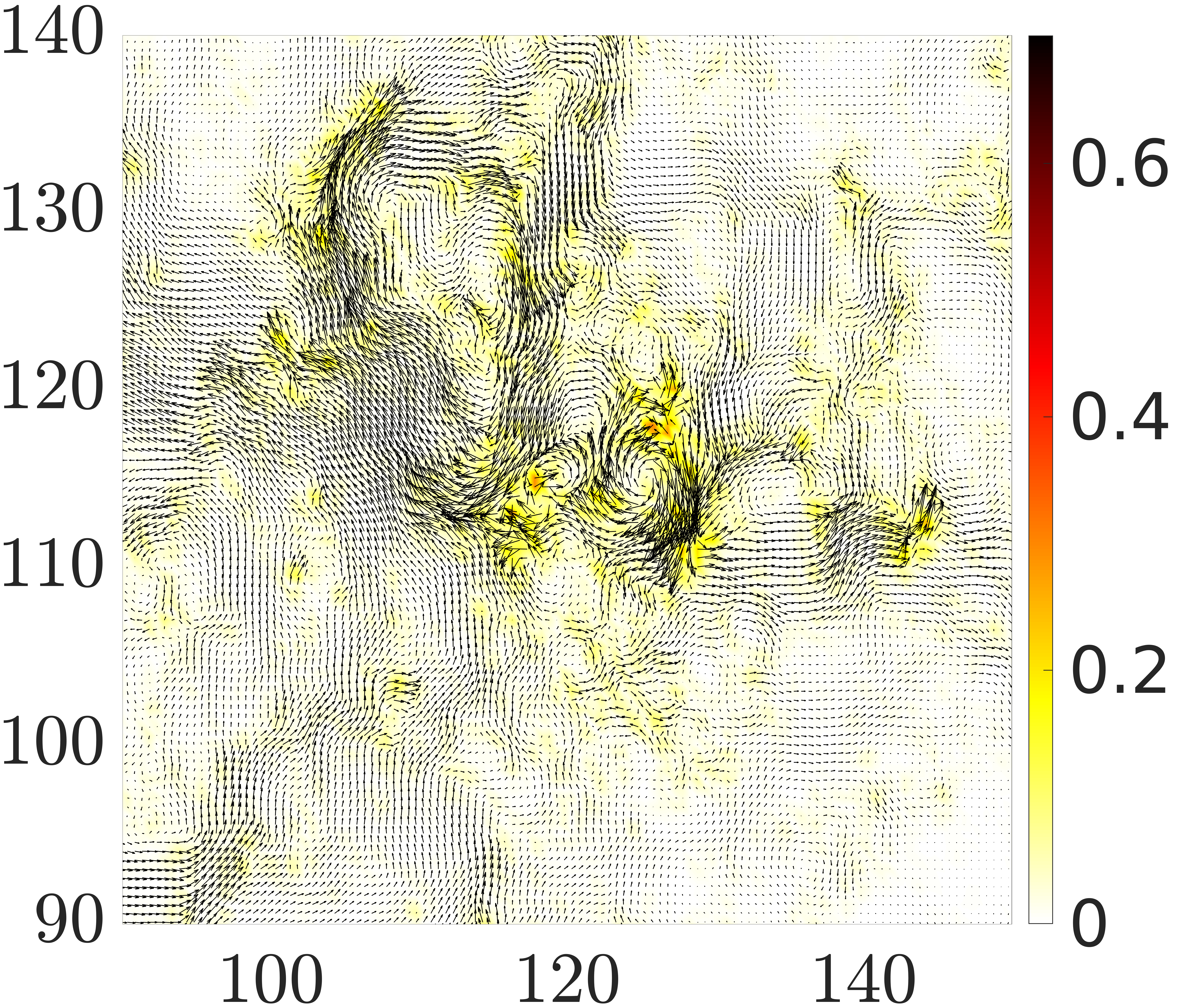}
	\caption{Heat map of the dipole field  $\C P^\alpha\equiv \partial_\beta Q^{\alpha\beta}$ for $P_0=4.5$, in the window of Fig.~\ref{Q}.}
	\label{dip}
\end{figure}

A typical map of the quadrupolar fields computed in this fashion, with the arrows in the direction of the angle $\Theta$, are shown in Fig.~\ref{Q}. 
The arrows are pointing in the direction of the angle $\Theta$, note that here there is no preferred angle with respect to the principal stress axis \cite{12DHP,13DHP}.
Since the quadrupolar field is obviously non-uniform, we expect that its divergence would be quite important. Thus we compute the dipolar field $\B {\C P}$, as the latter is expected to be crucial for the way stress is distributed as a result of the plastic event. The dipolar field is simply computed as $\C P^\alpha\equiv \partial_\beta Q^{\alpha\beta}$ \cite{21LMMPRS,22MMPRSZ,22BMP,22KMPS,23CMPR}.
In Fig.~\ref{dip} we present the divergence of the quadruopolar field $\B Q$ that is shown in lower panel of Fig.~\ref{Q}. At this point the important observation is that this field is not zero.  
The implications for modeling amorphous solids under shear are deferreed to the last section.

\subsection{Transition for varying pressure}
\label{transition}
The main question that we raise here is whether there exists
a clear transition, as a function of an intensive parameter in a given athermal amorphous system,
separating a regime in which the mechanical response tends to jump from Eq.~(\ref{renelas}) to Eq.~(\ref{amazing}) with a finite value of $\kappa$. We show next that in 2-dimensions the answer is affirmative, the intensive parameter for a granular jammed system is the pressure, and the transition is indeed clear.  

\begin{figure}
	\includegraphics[width=1.0\linewidth]{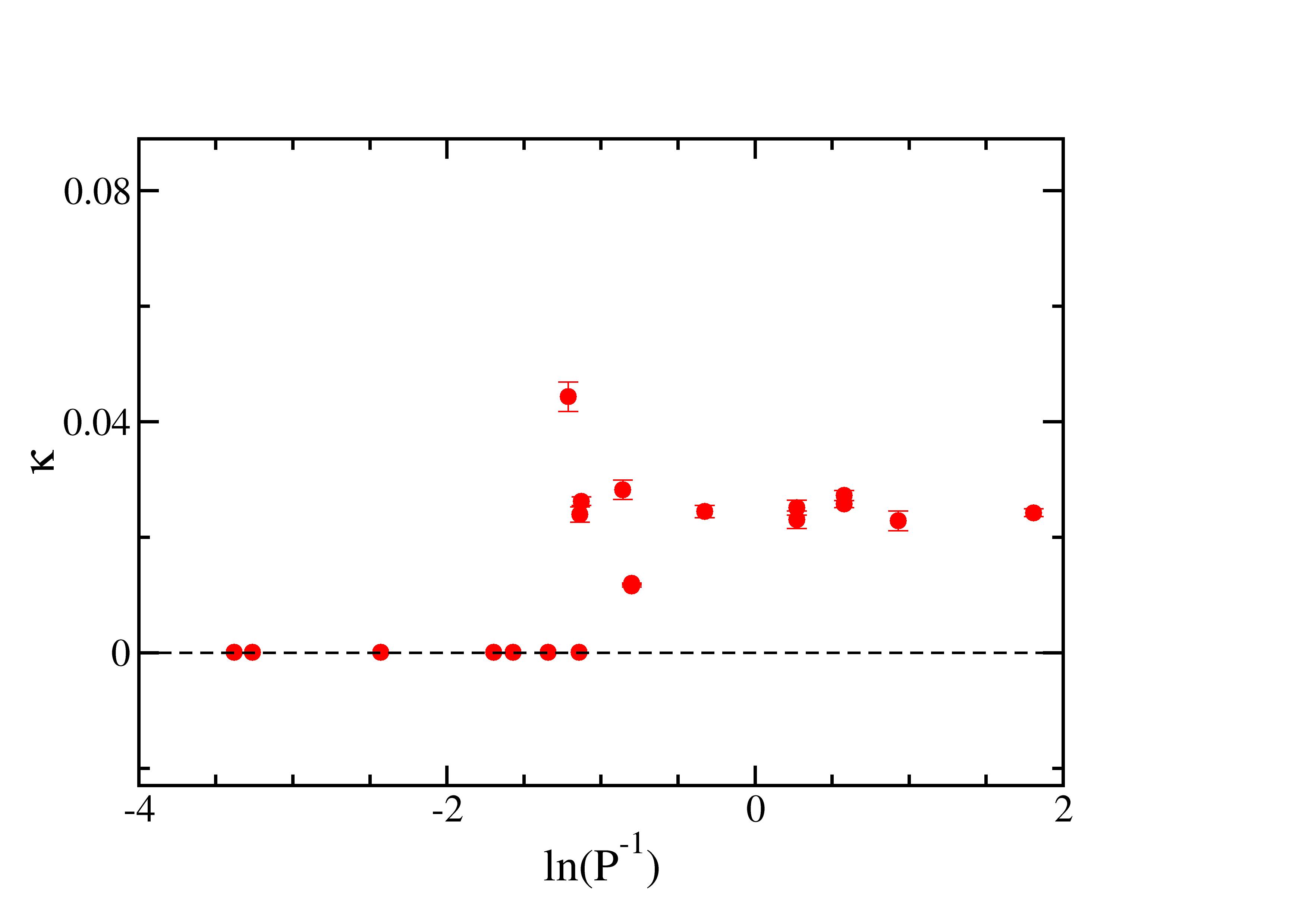}
	\caption{The screening parameter $\kappa$ as a function of the logarithm of the inverse pressure. A transition between material phases with quasi-elastic response and with anomalous response is clearly observed, with $\kappa\approx 0.023$.}
	\label{kapP}
\end{figure}

To demonstrate the transition we investigate frictionless assemblies of small disks that are at mechanical equilibrium,  prepared with a desired target pressure $P$ and confined in two-dimensional annulus as discussed in Sect.~\ref{radinf}. Details of the simulations can be found in Ref.~\cite{23JPS}. After achieving a mechanically stable configuration at a target pressure, we inflate the central disk by a desired percentage.  The displacement field exhibits qualitatively different appearance at high and low pressures.
At high pressures the displacement field is centered around the inflated disk as is expected from Eq.~(\ref{renelas}). In contrast, at low pressure the displacement field is spread out throughout the system, in correspondence with Eq.~(\ref{amazing}). This spread is due to plastic events that span the system. We refer to these as ``avalanches".
A quantitative comparison is provided by plotting the radial component $d_r(r)$, cf. \cite{23JPS}.
The simulations indicate a clear transition from quasi-elastic to anomalous response. The best way to demonstrate the transition is to measure the screening parameter $\kappa$ as a function of the pressure. In Fig.~\ref{kapP} we present the measured screening parameter as a function of $\ln(P^{-1})$.
 For pressure $P\ge 3.5\pm 0.3$ the response is quasi-elastic with $\kappa=0$. For pressure $P\le 3.5\pm 0.3$ the response is anomalous. The scatter in the values of $\kappa$  in the anomalous regime is typical to the considerable sample-to-sample fluctuations in the values of the screening parameter. We should note that once the screening parameter differs from zero it appears quite independent of pressure. 
 
 The next challenge is to predict at which pressure the transition should by observed, and what is the expected screening parameter $\kappa$. Both questions receive detailed attention in Ref. \cite{23JPS}, with constructive answers obtained using the scaling theory of the jamming transition. The reader is referred to that paper for further details. 

\section{The road ahead}

The short review presented above should whet the reader's appetite for more. We note that the novel physics that was revealed regarding the role of screening in the mechanical response of amorphous solids was based on linear theory (Lagrangians expanded to quadratic order), and for time independent responses to strains at the boundaries. Extending the theory to nonlinear order requires care. One could follow ideas of classical nonlinear elasticity, but these were shown to be fraught with difficulties for amorphous solids \cite{11HKLP}. Alternatively, one can seek nonlinear terms in the quadrupole and dipole interaction. Such a strategy was shown to be useful in the context of meatamaterials \cite{20BLBM}. The quasi-static mechanics of amorphous solids includes interesting instabilities like shear banding, and time dependent strains reveal phenomena like shear thickening or shear thinning. The extension of the theory presented above to include nonlinearities and time dependence is work in progress, to be discussed in future publications. In addition, the role of topological charges in materials other than amorphous solids, like metamaterials and active materials, is a subject of great interest, and a promising field of development for ideas of the type discussed above.  Finally, one needs to rethink how to model the response of amorphous solids under stress. The literature offers a number of `elasto-plastic' models, but these assume that stress drops are distributed in space according to the linear elastic Eshelby kernel. As we see from the present review, this assumption has to be revisited, and probably revised. In short, a lot of useful work lies ahead, and will hopefully lead to a better understanding of the physics of amorphous solids. 


\bibliography{ALL.anomalous}

\end{document}